\documentclass[conference]{IEEEtran}
\IEEEoverridecommandlockouts
\usepackage{cite}
\usepackage{amsmath,amssymb,amsfonts}
\usepackage{algorithmic}
\usepackage{graphicx}
\usepackage{textcomp}
\usepackage{xcolor}
\usepackage{flushend}
\def\BibTeX{{\rm B\kern-.05em{\sc i\kern-.025em b}\kern-.08em
    T\kern-.1667em\lower.7ex\hbox{E}\kern-.125emX}}

\newcommand*\fourteenpt{\fontsize{14}{15.5}\selectfont\textrm}
\renewcommand{\figurename}{Figure}

\usepackage{svg}

\begin{document}

\title{\textbf{\fourteenpt{ Towards a Methodology for the Development of Routing Algorithms in Opportunistic Networks}}}

\author{
\IEEEauthorblockN{Diego Freire}
\IEEEauthorblockA{\textit{Dept. of Information and Communications Engineering}\\
\textit{Universitat Autònoma de Barcelona}\\
Barcelona, Spain \\
email: diego.freire@deic.uab.cat}
\and
\IEEEauthorblockN{Sergi Robles}
\IEEEauthorblockA{\textit{Dept. of Information and Communications Engineering} \\
\textit{Universitat Autònoma de Barcelona}\\
Barcelona, Spain \\
email: sergi.robles@uab.cat}
\and

\IEEEauthorblockN{\centerline{Carlos Borrego}}
\IEEEauthorblockA{\centerline{\textit{Dept of Mathematics and Computer Science}} \\
\textit{\centerline{Universitat de Barcelona}}\\
\centerline{Barcelona, Spain} \\
\centerline{email: carlos.borrego@ub.edu}}

}

\maketitle


\begin{abstract}
 This paper introduces a methodology for the development of routing algorithms that takes into consideration opportunistic  networking. The proposal focus on the rationale behind the methodology, and highlights its most important stages and  components. It also discusses the importance of two core elements in the process of  protocol  designing: the scenario selection, based on essential characteristics, and the choice of standard evaluation metrics. As of now, there has been no common methodology for developing new routing algorithms, and this has led to proposals difficult to compare, to evaluate, and lacking a rigorous objectivity ensuring fairness. Thus, there is the urgent need to propose, agree, and use a common methodology for the development of routing algorithms. 
\end{abstract}
\renewcommand\IEEEkeywordsname{Keywords}


\section{Introduction}
The efficiency and performance of a network depends completely on the routing algorithm. Nodes can be sparsely or densely distributed, there can be few or many messages, node buffers can be small or large, but at the end of the day, the responsibility of forwarding all messages from the source to the destination in the best way possible lies with the routing algorithm. Thus, the development of such protocols is of paramount importance for the sake of networks in general, and specially critical in challenged networks like opportunistic ones. In these last type of networks, nodes are irregularly distributed, not always accessible and message forwarding is only possible when there is a transient contact opportunity. In these conditions, the store, carry and forward strategy of Delay Tolerant Networking helps connecting the unconnected parts of the net.

Unfortunately, the process of developing new routing algorithms has not been in the focus of research, and this has led to disputable quality proposals, difficult to compare between them, and almost impossible to determine if they suit best for a given scenario. Although many proposals include simulated experiments repeated in several conditions, with different data sets, and even including very detailed network configurations, such as radio protocols, and interference models, they still lack the basic scientific approach allowing repeatability and comparison. It is true that many of these papers introduce the confrontation to other routing algorithms, but even in this case, the scenario selection and particular configuration is not guaranteed to observe, intentionally or not, a rigorous objectivity ensuring fairness. Moreover, few of these proposals present a final implementation showing its feasibility and allowing a realistic performance evaluation under real world conditions. 

Traditional networks have an end-to-end path available to transmit messages between nodes, but in Opportunistic Networks, this end-to-end path may never exist, delays and disruptions are part of the behaviour; therefore, opportunistic strategies makes communication possible. In the development of a routing algorithm, evaluation and testing are done by some assumptions (e.g., unlimited resources, limited resources, a limited number of messages, unlimited creation of messages, among others). These assumptions seek to recreate a real-world OppNet, but complexity and variability increases within each characteristic studied. 

Having seen this, there is the urgent need to propose, agree, and use a common methodology for the development of routing algorithms that also takes into consideration extreme scenarios, such as opportunistic networking. The process has to go from the basic idea for the routing strategy, to the mathematical analysis, model, simulation, software implementation of the algorithm, emulation, and finally the application of the routing strategy, testing real code in real scenarios. In this paper, we get grips with the problem, and introduce the basic rationale for such a methodology, highlighting its most important stages and components, and discussing the importance of two core elements in the process of protocol designing: scenario selection, based on essential characteristics, and standard evaluation metrics. We expect this methodology to faster the adoption of a common scientific approach to the development of new routing algorithms, and to give firm leverage in the production of high quality routing algorithms for OppNet.

The rest of the paper is structured as follows. Section \ref{sec:relatedwork} introduces the state of the art on opportunistic networks, evaluation strategies and development methodologies. Section \ref{sec:methodology} presents the methodology in our proposal. Then, scenarios and metrics are shown in Section \ref{sec:scenarios}. Finally, Section \ref{sec:discusion} discusses our contributions and implications. 


\section{Related Work}\label{sec:relatedwork}

In this Section, we study the state of the art of the development of routing algorithms in the opportunistic networks field, the evaluation strategies and methodology used in development of new opportunistic routing algorithms
\subsection{Opportunistic networks}

An opportunistic network\cite{b1}, also known as OppNets\cite{b20},  is a set of mobile devices commonly called nodes that exchange information between them exploiting direct communication opportunities to perform an end-to-end transfer of data. Nodes communicate with each other even if an end-to-end route never exist\cite{b11}. Furthermore, nodes are not supposed to possess or acquire any knowledge about the network topology.  
With the growth of the use of mobile devices in recent years, opportunistic networks have become a significant field of research. Opportunistic networks allow a flexible and highly dynamic connection between the nodes\cite{b2}. Any node can join or leave the network at any time. \\

The applications of opportunistic networks\cite{b12} are cellular network offloading, communication in challenged areas, censorship circumvention and proximity-based applications and Internet of Things (IoT)\cite{b13}, among others.  

Topology network is continuously changing due to the constant movement of the nodes, and the communication routes between senders and receivers are neither direct nor static. This communication capability allows the use of opportunistic networks in new applications. Before opportunistic networks, applications based their operation on an end-to-end connection path, however, when such connection is not possible, opportunistic networks present a solution, since the information is sent "opportunistically" hop by hop between source to destination using the "Store-Carry-and-Forward" approach \cite{b3}.

\subsection{Evaluation strategies in opportunistic routing algorithms}

A standard methodology that allows the evaluation of routing algorithms in the OppNets field does not exist. Nevertheless, the nonexistence of a comparing method does not mean that a comparison is not possible. 

One of the main ways of evaluating opportunistic routing algorithms is measuring their performance when sending information from a source to a destination. Some authors model message dissemination performance by analysing first the behaviour of the OppNet when some characteristics of the network vary, such as density, size of the messages, duration of the contacts, etc. \cite{b4}.\\

Regarding the use of datasets for network behaviour simulation, and according to \cite{b5}, most authors use several common scenarios like Haggle\cite{b18}, MIT\cite{b17} or Cambridge\cite{b16}. Finding good scenarios to evaluate routing algorithms is not easy. The elevated cost of deploying real test-beds and the non-existence of a suitable simulator accounting for all real characteristics make it really difficult to find adequate traces to perform realistic simulations \cite{b5}. 


\subsection{Methodology in the development of new opportunistic routing algorithms}

A small part of the research community that works on challenged networks, such as opportunistic networks has pointed out the necessity of finding new methodologies in the development of new opportunistic routing algorithms. Their concern is focused on involving the engineering process in all stages between an original network proposal and its validation in real applications. The problem is that a lot of networks research proposals in this context rely solely on simulations to validate the proposed protocols without going any further. 

The authors of studies like \cite{b7}, draw attention to the fact that network decisions, such as routing or delivery ones are rarely implemented in real network platforms. Additionally, if they are, the validation of the proposed code is usually performed at a very small scale. That is why, there has been an enormous effort from this part of the research community on developing new emulation platforms to help with this problem. By using emulation tools, demanding scenarios can be tested and provide a  a lightweight emulation solution that bridges the gap between pure simulation and real-world experimentation.


\section{Methodology for developing routing algorithms}\label{sec:methodology} 

Using a sensible, sound methodology is indispensable to get good routing algorithms. This methodology has to observe the basic scientific method, allowing repeatability, fair comparison, and common scenario representation. In this Section we propose the basic steps of the process of developing new routing algorithms fulfilling the aforementioned requirements, and discuss about the selection of the test scenarios and performance metrics for comparison.

\subsection{Methodology stages}

The methodology has seven well differentiated stages, as shown in Figure \ref{fig:methodology}:

\begin{figure}
    \centering
    \includegraphics[width=7cm]{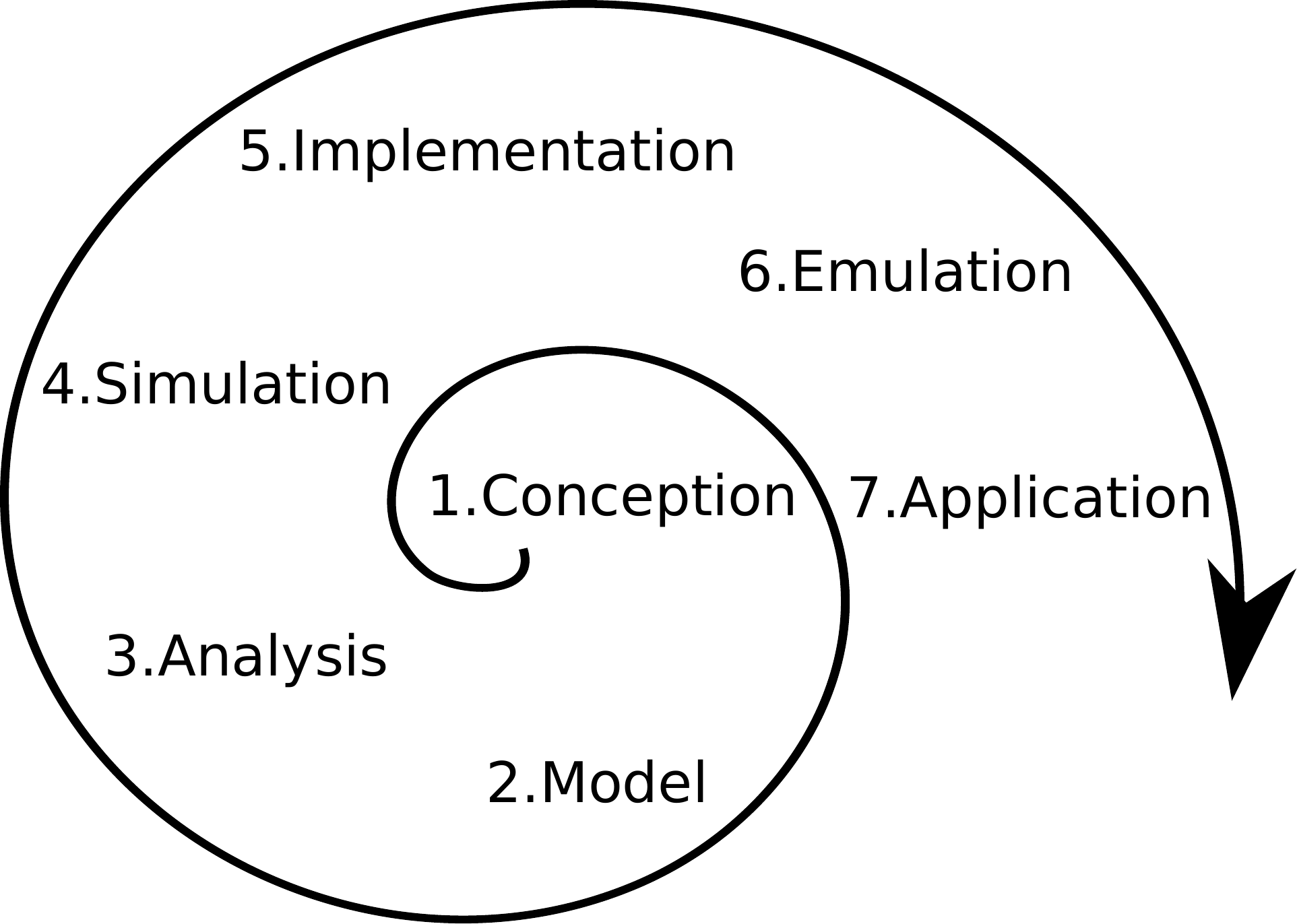}
    \caption{Methodology for developing routing algorithms consisting of seven stages.}
    \label{fig:methodology}
\end{figure}

\subsubsection{Conception} First it comes the initial idea behind the algorithm, the conception of the mechanisms. This is normally triggered by some essential feature of the scenario, such as high node density, or by a particular theory, such as the history of encounter of nodes and its transitivity. 

\subsubsection{Model} After the initial stage of conception, the idea must be reified into a particular mathematical model, which can then be analyzed formally.

\subsubsection{Analysis} Once the proposal is modeled, it can be analysed. During this analysis, the mechanisms and procedures can be checked, some theoretical results can be obtained, and basic limitations can be identified. 

\subsubsection{Simulation} The next stage is simulation. In a simulator, the model can be tested in a given set of scenarios. Eventhough these scenarios involve datasets that come from the real world (e.g., real traces from vehicles or people), or even if the simulator simulates very accurately all  network protocols involved, the model under evaluation is usually executed based on pseudo-code. This does not prove that the system being designed can eventually be deployed and used for real. Results obtained through simulation can be deceptive, creating a misleading feeling of scientific correctness. Indeed, as observed in \cite{b10}, the credibility of simulation results tends to decrease as the use of simulation increases.

\subsubsection{Implementation} The final validation of a routing algorithm should always be based on real full-featured code (accounting for example for memory management or concurrency issues), rather than on the pseudo-code used in simulations. In this stage, a code is produced so that the algorithm can be used on a real scenario. The implementation itself shows the feasibility of the algorithm.

\subsubsection{Emulation} Testing real code in real conditions can be difficult and tricky, especially when these situations may involve the mobility of hundreds of nodes during hundreds of hours. Emulation is an approach that helps with this respect, allowing to run real code in tightly controlled (and repeatable) conditions. This stage is the link between a proof-of-concept implementation and the deployment of a software that is useful in the real world and behaves as predicted. 

\subsubsection{Application} The last stage of this methodology is testing the routing algorithm in a real environment, with real devices and users. This is the ultimate test that shows how the designed algorithm behaves in the real world and allows to evaluate.

The application of the methodology should not be strictly sequential. Some of the work in one stage can help to improve some of the previous stages. For example, the results of the analysis can help modifying the model to take into account a new variable, or the emulation results can help to detect and correct bugs of the implementation.  

Applying a methodology like the one described is necessary, but not enough to produce good quality routing algorithms. There are two elements that have also to be considered: the scenarios, and the performance metrics. The simulation and emulation stages need some scenarios, including the position of nodes during a time window and the messages that are sent along with other information. This is important for two reasons. In the first place, these scenarios have to be public to reproduce the results at any time, and to fairly compare different routing algorithms in the same conditions. Secondly, the scenarios need to be representative of the real environments the routing algorithm is going to be used in. The second element to be considered is the performance metrics of the algorithm. Again, there are two main reasons for this. The first one is that to evaluate how good is a routing algorithm for a given scenario, there have to be some evaluation functions. These functions, or metrics, will tell how the network performs when this routing algorithm is in operation, and thus can determine for which scenarios it is more appropriate. The second reason is that these metrics allow a direct comparison to other protocols for the same scenario. 


\section{Scenarios and metrics in opportunistic networks}\label{sec:scenarios}

Different scenarios and metrics are generally used in articles to measure the performance of Routing Algorithms. As it is difficult to reproduce real-world conditions, the use of scenarios tries to create a model of them, assuming performance will be similar. To be in the safe side, many papers use several scenarios to show the algorithm has a large scope of applicability. However, articles use to pay little or no attention to the selection and definition of these scenarios, neglecting the importance they deserve in the significance of the results.


\subsection{Components}

Scenarios of OppNet consist of a set of nodes and their positions during a time frame. When analysing the behaviour of a routing algorithm on a scenario, more details have to be provided, such as a set of messages (with recipients and size), and the communication range of the nodes.

Nodes in an OppNet normally communicate wirelessly. Nodes can receive, drop, store, carry and forward messages. When a node receives a message, the decision whether a message must be stored, carried, dropped or forwarded is made by the routing algorithm. The routing algorithm makes the decision of which nodes a message is forwarded to. 

As we talk before scenarios are a representation of real-world, therefore their  characteristics must help to reproduce certain real-world behaviour.
Among others, the characteristics of a scenario may include the set of positions, granularity, node range, node density, and buffer size. For the sake of simplicity, we can consider the scenario as the set of nodes, the set of messages and the contacts between nodes:

\begin{equation}
S=(Nodes, Messages, NodeContacts)\label{eq1}
\end{equation}


\subsection{Selection of scenarios}
As we have seen, scenarios play a very important role for the development of routing algorithms. As diversity is a key factor to guarantee representation of real world applications, different sources of scenarios have to be considered. They can be synthetically created, which allows to force some scenario characteristics, like a given node density. They can also come from real world traces captured in live situations. As suggested by Kotz et al. in \cite{b15}, create a new scenario is expensive and challenging. The Community Resource for Archiving Wireless Data At Dartmouth (CRAWDAD) allows sharing data sets across the scientific community. The real-world data help us to understand the behaviour of real users. 
Common well-known scenarios already used in literature\cite{b21}, like Cambridge, Info5, Taxis, MIT or Haggle could also be used as possible scenarios. Within the many scenarios that have different characteristics, scenario selection is a crucial part of the development of a new opportunistic routing algorithm. The output performance of some scenarios is the same, even that those scenarios do not share characteristics. 

Because it is impossible to test the routing algorithms in all possible scenarios, making a selection of scenarios is required. This process have to be scientifically justified, to be representative enough and avoid any possible bias. Just having different traces that generate different results without analyzing the entire spectrum of action of the OppNet comprehensively would result insufficient. These fine selection of scenarios will act as a representation of the whole scope of opportunistic networks.

\subsection{All-in-One scenario trap}
A valuable scenario aims to be a good representation of reality. This representation must introduce as many elements as the real event contains, but this could end up into an unrealistic task due to the number of characteristics involved. Our research found more than seventy characteristics used as tuning settings of the so-called scenarios. Given those seventy characteristics, even limiting the operativity of each characteristic as binary,  the number of the scenarios is unrealistic to manage. 
Therefore, modelling a scenario requires a balance between simplification and real-world accuracy has a direct implication of usability. Not every characteristic must be taken into account. Those characteristics that are not involved in the scenarios are going to be present in the other phases of the routing algorithm development.  Oversimplification of characteristics could lead to a useless representation of the phenomena, and the results are not useful.
Our proposal claims that instead of build an "All-in-One" scenario, the development of a set of different scenarios, where those scenarios must give different performance results with the same algorithms.

\subsection{Performance metrics}
In a scenario, messages have to be delivered from the origin to the destination taking advantage of the communication opportunities. To achieve this, in this process, several copies of the same message are generated. Messages can be successfully delivered, they can be dropped, for example if there is not enough buffer for them, or they may not reach their final destination, for example because their life time is over. A metric is a function that gives the measure of a certain property of a given scenario, such as message drop, latency, node inter-contact time, delivery ratio, overhead ratio, delivery cost, average number of hops, wastage index or average delay, among others. 

The evaluation function of an algorithm in a scenario provides the set of some metric measurements for that scenario, as shown in \eqref{eq2}. The result of this evaluation is multidimensional, for having just one number to compare different algorithms does not allow an accurate comparison. An algorithm may be better for a specific metric, but worse for another one. It is the final application that will determine which algorithm is the most appropriate, and therefore it is convenient to preserve the whole set of metric results to have a better idea on how the algorithm behaves.

The evaluation function can be represented as:

\begin{equation*}
M={M_1, M_2, ..., M_n}
\end{equation*}

\vspace*{-0.5cm}

\begin{equation}
Eval(A, S, M)=( M_1(A, S), M_2(A, S) , ... , M_n(A, S) )\label{eq2}
\end{equation}

\noindent where $Eval$ is the evaluation function, $A$ is the routing algorithm, $S$ is the scenario, $M$ is the set of metric functions, and $M_i$ is the measurement of the metrics.

\medskip

Metrics should always refer to the same measurable properties, thus all proposals have to use exactly the same names for the metrics to avoid confusion.



\begin{figure}
    \centering
    \includegraphics[width=7cm]{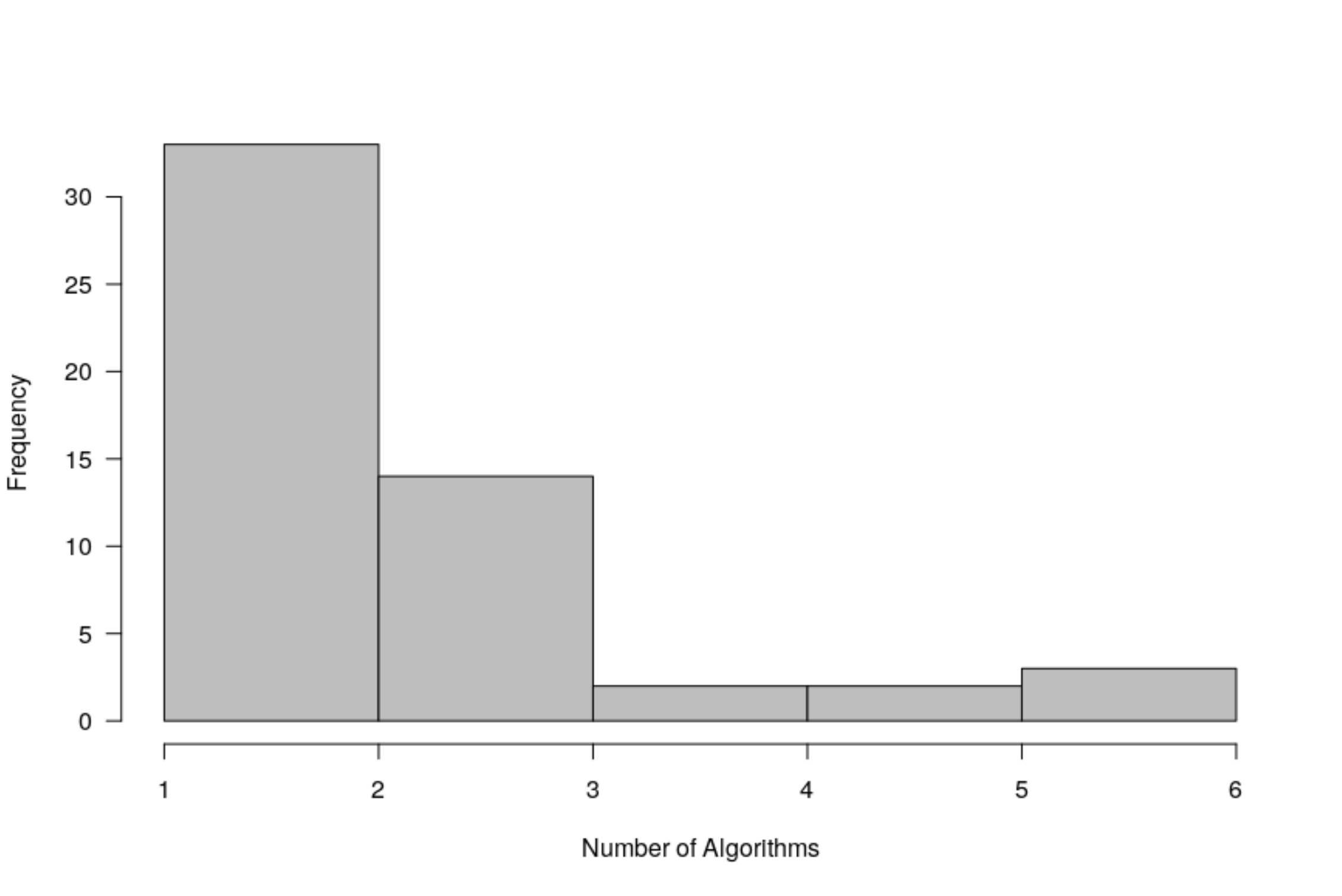}
    \caption{Number of compared-to routing algorithms.}
    \label{fig:comparison}
\end{figure}

\subsection{Routing comparison}

Our research shows how the performance comparison has been carried out so far when a new opportunistic routing algorithm has been presented; and how these practices could lead up to unfair comparisons. We studied more than 50 opportunistic routing algorithms. 

\begin{figure}
        \centering
        \includegraphics[width=0.5\textwidth, angle=0]{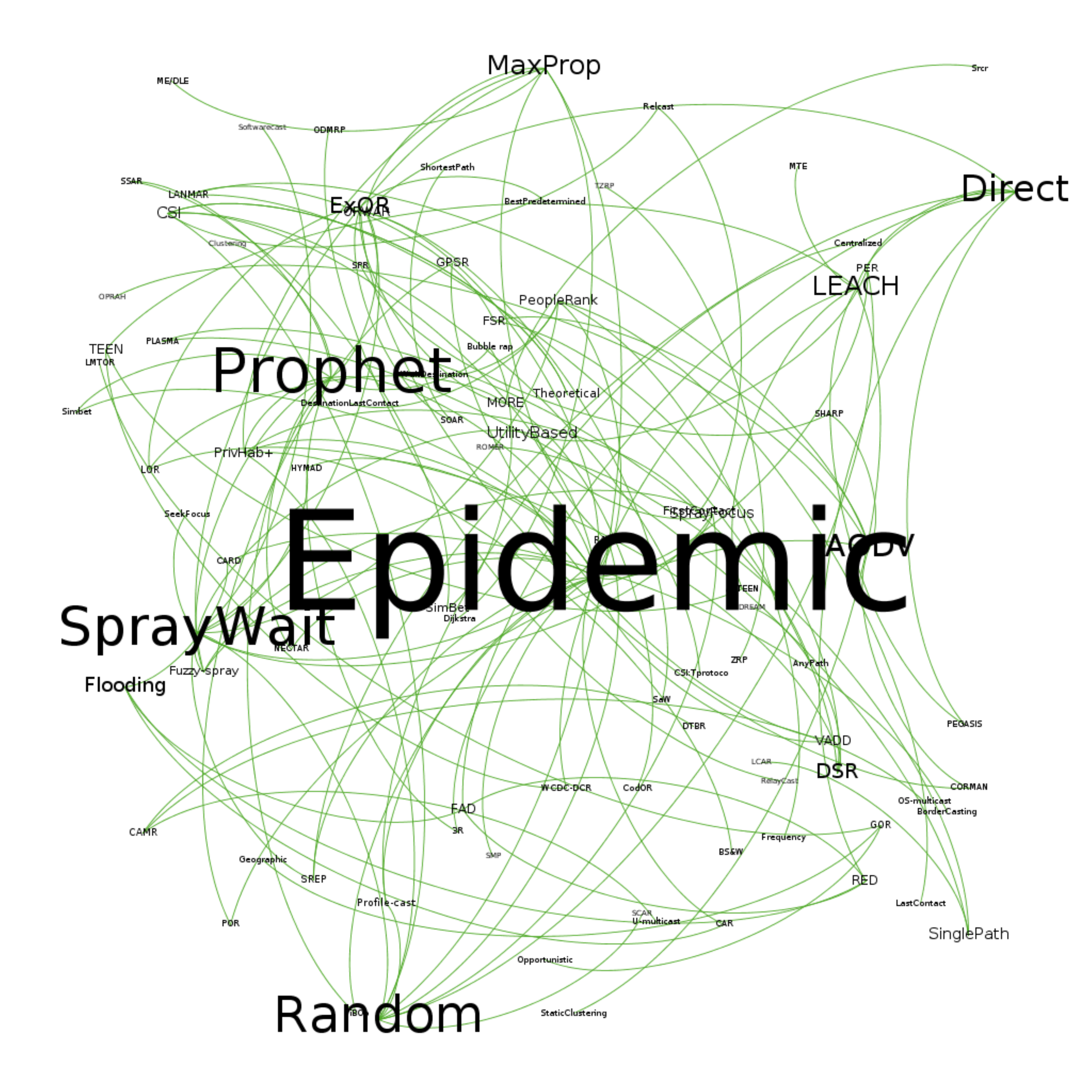}
        \caption{Cloud graph of routing algorithms comparison, edges indicates a direct comparison between a pair of routing algorithms in a paper}
        \label{fig:cloud}
\end{figure}

A comparison helps to evidence the improvement in the performance of a given task. Figure \ref{fig:comparison} shows the number of comparisons founded in the papers on routing algorithms in the literature. Then, Figure \ref{fig:cloud} is a cloud graph where edges indicate that the two routing algorithms connected are directly compared in some paper. This graph emphasizes the number of comparisons of a routing algorithm, The bigger the size of the font, the more times an algorithm has been compared to.

From the literature on routing algorithms, the information shown in Figure \ref{fig:comparison} and Figure \ref{fig:cloud} reaffirms our concern about a fair comparison.

We found that approximately a 62\% of the algorithms tests their performance with one or two algorithms when they are presented. On the other hand, less than 5\% of the reviewed algorithms present a comparison with 6 algorithms. No algorithm which is compared with more than 6 others has been found. 

In Epidemic routing, when a message needs to be routed from a source to a destination, the algorithm sends the message to all of their reachable neighbours. The algorithm does not have to make any decision whether to send a message or not. Having that in mind, the implementation of an Epidemic algorithm is not difficult at all. Figure \ref{fig:comparison} shows that most of the literature makes less than 2 comparisons and Figure \ref{fig:cloud} indicates the Epidemic routing is the most compared-to algorithm. That means that the comparison is centred around Epidemic routing and that routing algorithms are not been compared between them. Moreover, the few algorithms used in the comparisons are those that are implemented in traditional simulators. 
This is probably due to fact that the scientific community is paying more attention to the simplicity of the experimentation design rather than to the scientific soundness. 

Our approach is that every algorithm could be evaluated in a fairway. That means, be able to compare apples with apples.

The use common evaluation function (as defined in \eqref{eq2}) in the analysis of an algorithm in a scenario provides a deterministic outcome, thus avoiding intentional or unintentional bias. With a methodological performance evaluation, we can pick the best algorithms of our interest.

We know that opportunistic networks are a challenging field; therefore, the traditional way to develop routing algorithms is not enough to go from the idea to the application.



\section{Discussion}\label{sec:discusion}
When a new routing algorithm is proposed in the field of opportunistic networking, the proposal uses its own evaluation metrics and scenarios. Thus, even though these proposals normally compare their protocol to others, the testing environment, chosen scenarios, metrics and conditions do not guarantee that the comparison is totally fair. It is very easy to, unintentionally, create a bias favouring their own proposal just by selecting a scenario with some specific characteristics for which the algorithm has been designed to take advantage. In most cases, reproducing the results of a given proposal is practically impossible because the full scenario data is not publicly available, the way of applying the metrics is not completely clear, or simply because the implementations (or models of the protocol) are not given. As suggested by Bajpai et al. in \cite{b14}, the research on computer networks is more and more accepting research proposals that are non-reproducible as long as they appear plausible to the manuscript reviewers. We agree with this study on the importance and challenges on reproducibility.

This situation applies to routing algorithms in general, but is particularly notorious in the case of opportunistic networks, where there is more variety of scenarios and conditions. This does not facilitate at all the selection of a good routing algorithm for a certain application, for example, and the creation process of new protocols is as well weakened for no validated references are available to fairly compare them to other proposals.   

A way of alleviating these effects and giving better prospects to the (useful) development of new routing algorithms in ooportunistic networking is by using a methodology like the one presented in this paper. This can decisively help on the difficult task of starting a cultural change on routing algorithm development, proposal evaluation, and algorithm selection.


\section{Conclusions}\label{sec:conclusions}
In this paper, we have presented a methodology for the development of routing algorithms in Opportunistic Networking based on seven stages. Following this methodology, the development of new routing algorithms will improve , increasing their quality and fair comparison to others.  Additionally, we have discussed the importance of two core elements in the process of protocol designing: the scenario selection and the choice of standard evaluation metrics. 




\section*{Acknowledgment}

This work is partly funded by Secretaria de Educación Superior, Ciencia, Tecnología e Innovación (SENESCYT, ECUADOR), by the Catalan AGAUR 2017SGR-463 project, and by the Spanish Ministry of Science and Innovation TIN2017-87211-R project.




\end{document}